\documentclass[astrosymb,twocolumn,floatfix]{aastex631}
\usepackage{physics}
\usepackage{graphicx}

\newcommand{\ICTS}{\affiliation{International Centre for Theoretical Science, Tata Institute of Fundamental Research, Bangalore - 560089, India}}
\newcommand{\UChicago}{\affiliation{Department of Physics, The University of Chicago, 5640 South Ellis Avenue, Chicago, Illinois 60637, USA}}
\newcommand{\IISER}{\affiliation{Department of Physics, Indian Institute of Science Education and Research Pune, Maharashtra, 411008, India}}
\newcommand{\CITA}{\affiliation{Canadian Institute for Theoretical Astrophysics, University of Toronto, 60 St George St,  Toronto, ON M5S 3H8, Canada}}
\newcommand{\UofT}{\affiliation{David A. Dunlap Department of Astronomy and Astrophysics, and Department of Physics, 60 St George St, University of Toronto, Toronto, ON M5S 3H8, Canada}}
\newcommand{\KICP}{\affiliation{Kavli Institute for Cosmological Physics, The University of Chicago, 5640 South Ellis Avenue, Chicago, Illinois 60637, USA}}
\newcommand{\EFI}{\affiliation{Enrico Fermi Institute, The University of Chicago, 933 East 56th Street, Chicago, Illinois 60637, USA}}
\newcommand{\UChicagoAA}{\affiliation{Department of Astronomy and Astrophysics, The University of Chicago, 5640 South Ellis Avenue, Chicago, Illinois 60637, USA}}

\newcommand{\rev}[1]{\textbf{}}

\begin{document}
	
	\title{Inferring host-galaxy properties of LIGO-Virgo-KAGRA's black holes}

	\author[0000-0002-4103-0666]{Aditya Vijaykumar}
	\CITA
	\ICTS
	\UChicago
	
	\author[0000-0002-1980-5293]{Maya Fishbach}
	\CITA
	\UofT
	
	\author[0000-0002-0298-4432]{Susmita Adhikari}
	\IISER
	
	\author[0000-0002-0175-5064]{Daniel E. Holz}
	\UChicago
	\EFI
	\KICP
	\UChicagoAA
	
	\begin{abstract}
		
		Observations of gravitational waves from binary black hole (BBH) mergers have measured the redshift evolution of the BBH merger rate. 
		The number density of galaxies in the Universe evolves differently with redshift based on their physical properties, such as their stellar masses and star formation rates. 
		In this work we show that the measured population-level redshift distribution of BBHs sheds light on the properties of their probable host-galaxies. 
		We first assume that the hosts of BBHs can be described by a mixture model of galaxies weighted by  stellar mass or star formation rate, and find that we can place upper limits on the fraction of mergers coming from a stellar mass weighted sample of galaxies.
		We then constrain parameters of a physically motivated power-law delay-time distribution using GWTC-3 data, and self-consistently track galaxies in the \textsc{UniverseMachine} simulations with this delay-time model to infer the probable host-galaxies of BBHs over a range of redshifts. We find that the inferred host-galaxy distribution at redshift $z=0.21$ has a median star formation rate $\sim 0.9\,M_\odot\mathrm{yr}^{-1}$ and a median stellar mass of $\sim 1.9 \times 10^{10}\,M_\odot$. We also provide distributions for the mean stellar age, halo mass, halo radius, peculiar velocity, and large scale bias associated with the host-galaxies, as well as their absolute magnitudes in the B- and ${ \rm K_s}$-bands. 
		Our results can be used to design optimal electromagnetic follow-up strategies for BBHs, and also to aid the measurement of cosmological parameters using the statistical dark siren method.
	\end{abstract}

	\section{Introduction} \label{sec:intro}

	host-galaxies of astrophysical electromagnetic (EM) transients have offered interesting insights into the astrophysics of their progenitors.
	For instance, the host-galaxy population of Type Ia supernovae and core collapse supernovae are different (e.g. \citealt{2013ApJ...778..167F}); this is expected given that both these classes have different progenitors. 
	Similarly, long and short gamma ray bursts (GRBs)  are known to trace different host-galaxy properties~\citep{2010ApJ...708....9F, 2014PASP..126....1L, 2017MNRAS.467.1795L, 2022ApJ...940...56F,2022ApJ...940...57N}. 
	The inferred star formation histories of short GRB host-galaxies have also been used to constrain their delay-time distribution, finding a preference for low delay times but also evidence for a high delay-time tail~\citep{2022ApJ...940L..18Z}. 
	Interesting conclusions about fast radio burst (FRB) progenitors have also been drawn using their host-galaxies~\citep{2022AJ....163...69B}.
	The follow-up and cataloging of host-galaxies of supernovae, short GRBs, and FRBs continues to be an important research direction. 
	
	In a similar vein, information about the host-galaxies of gravitational-wave (GW) transients could shed light on many interesting aspects of their formation and evolution.
	For instance, \cite{2019ApJ...878L..12S} propose to use the population of host-galaxies of $\order{1000}$ binary neutron stars (BNSs) to measure their underlying delay-time distribution, whereas \cite{Adhikari:2020wpn} find that $\order{10}$ detections of localized BNSs would be sufficient to provide interesting constraints on BNS progenitor models, including the relation to star formation and delay-time distributions.
	Any information about probable host-galaxies could also be used to reduce the number of candidate galaxies searched to localize a GW transient, and in designing optimal weighting schemes in the statistical method to measure cosmological parameters~\citep{1986Natur.323..310S, 2012PhRvD..86d3011D, 2018Natur.562..545C, 2019ApJ...871L..13F, 2021ApJ...909..218A, 2023ApJ...949...76A}.
	
	A number of works have made assumptions on the formation and evolution of both galaxies and compact binaries to forward-model the distribution of GW transient host-galaxies~\citep{OShaughnessy:2009szr, Lamberts:2016txh, Mapelli:2019bnp, Toffano:2019ekp, Artale:2019doq, Santoliquido:2022kyu,Srinivasan:2023vaa, 2023MNRAS.523.5719R}. For example, \cite{Santoliquido:2022kyu} found that, contingent on their modeling assumptions, binary black holes (BBHs) prefer merging in high mass galaxies. 
	However, unlike their EM cousins, host-galaxies of GW transients cannot usually be identified, owing to poor pointing accuracy of GW detectors. The notable exception to this is the binary neutron star (BNS) merger GW170817~\citep{2017PhRvL.119p1101A}, whose electromagnetic emission helped localize the source to the galaxy NGC 4993~\citep{2017ApJ...848L..12A}. 
	It comes as no surprise that although the LIGO-Virgo-KAGRA (LVK)~\citep{2015CQGra..32g4001L, 2015CQGra..32b4001A, 2021PTEP.2021eA101A} network of detectors have reported the detection of $\sim 90$ BBHs \citep{2021arXiv211103606T}, none of them have been confidently associated with a host-galaxy\footnote{There is a claimed electromagnetic counterpart ZTF19abanrhr~\citep{2020PhRvL.124y1102G} to the massive BBH GW190521~\citep{2020PhRvL.125j1102A}, believed to be produced due to the interaction of the BBH with the accretion disk of an active galactic nucleus; however, there is insufficient evidence to confidently associate GW190521 with ZTF19abanrhr~\citep{2021ApJ...914L..34P,2021CQGra..38w5004A}. }. 
	Even with upgrades to the current detector network with the inclusion of LIGO-India~\citep{2022CQGra..39b5004S}, or with more sensitive next generation detectors such as Cosmic Explorer~\citep{2021arXiv210909882E} and Einstein Telescope~\citep{2010CQGra..27h4007P}, only a small fraction of the BBHs are expected to be sufficiently localized such that the host-galaxy can be confidently identified~\citep{Nishizawa:2016ood, Chen:2016tys, Borhanian:2020vyr, Branchesi:2023mws, 2023arXiv230710421G,2020LRR....23....3A}. 
	
	Thinking about the detected BBHs as a population instead of individual events have allowed measurements of the underlying mass and spin distribution of black holes in binaries, as well the merger rate of binaries as a function of redshift, $z$~\citep{2019ApJ...882L..24A, 2021ApJ...913L...7A, 2023PhRvX..13a1048A}.
	Specifically, the redshift-evolution of the merger rate~\citep{2018ApJ...863L..41F,2021ApJ...912...98F,2021ApJ...913L...7A,2023PhRvX..13a1048A} has yielded constraints on the delay-time distribution and formation metallicity of BBHs~\citep{2021ApJ...914L..30F,2023arXiv230715824F,2023arXiv231017625T}, and properties of globular clusters~\citep{2023MNRAS.522.5546F}. 
	In this work, we show that the redshift evolution of the merger rate inferred from the latest GW transient catalog GWTC-3~\citep{2021arXiv211103606T} can already begin to constrain the host-galaxy parameter space of detected BBHs. This results directly from the observation that galaxies weighted by different physical properties have different evolutions for their (weighted) number density (see e.g.~\citealt{2007ApJ...665..265F}). For instance, there are more star-forming galaxies per unit volume at $z=1$ as compared to $z\approx 0$, while there are a lesser number of massive galaxies at higher redshifts. 
	
	Our work is structured as follows. In Section~\ref{sec:sm_or_sfr} we assume that the measured GWTC-3 merger rate is described by a simple mixture model of galaxies weighted by their star formation rate and stellar mass, and infer the mixing fraction between them. We derive an upper limit for the possible contribution of BBHs from galaxies weighted by stellar mass.
	In Section~\ref{sec:physical_constraints} we constrain the delay-time distribution of BBHs using a physically-motivated model. We also seed the star-formation histories of galaxies taken from \textsc{UniverseMachine}~\citep{Behroozi:2019kql} with the inferred delay-time distribution to study the resulting host-galaxy properties, and present results derived using this procedure. We summarize our results and dwell on directions for future work in Section~\ref{sec:summary}. Throughout we assume cosmological parameters consistent with the Planck 2018 data release~\citep{2020A&A...641A...6P} as implemented in the \texttt{Planck18} class in \texttt{astropy}\footnote{We note that the \textsc{UniverseMachine} simulation snapshots use a slightly different set of cosmological parameter values ($h=0.678$, $\Omega_m=0.307$) compared to the \texttt{Planck18} implementation in \texttt{astropy} ($h=0.6766$, $\Omega_m=0.30966$). Both of these estimates are consistent with \citet{2020A&A...641A...6P}; we do not expect this small discrepancy to impact our results significantly.}~\citep{2013A&A...558A..33A,2018AJ....156..123A}, and that galaxy formation histories are well-emulated by \textsc{UniverseMachine}.
	
	\section{Do BBHs follow stellar mass or star-formation rate?}
	\label{sec:sm_or_sfr}
	
	We first demonstrate that it is possible to differentiate between host-galaxy populations of BBHs using the redshift evolution of the merger rate. If BBHs were hosted by a sample of host-galaxies purely weighted by their star formation rate (SFR), the rate of BBHs would increase as a function of redshift, roughly as $(1 + z)^{2.5}$, up to $z\sim 1.5$,  following the evolution of the cosmic star formation rate density (cSFRD)\footnote{\rev{Throughout this paper, SFR refers to the star-formation rate of an individual galaxy, whereas cSFRD refers to the SFR integrated over all galaxies per unit volume as a function of redshift.}} \citep{Behroozi:2019kql, Madau:2014bja, Madau:2016jbv}\footnote{We use the cSFRD inferred by \textsc{UniverseMachine}~\citep{Behroozi:2019kql} instead of other prescriptions~\citep{ Madau:2014bja, Madau:2016jbv} to be consistent with choices made later in this work.} which peaks at that redshift\footnote{This is a good assumption for current GW detectors, whose horizon redshift ($z\sim1$) is well below the peak star-forming epoch.}. On the other hand, if BBHs were hosted by a sample of galaxies purely weighted by their stellar mass, $M_*$, the BBH number density would decrease as a function of redshift; this is because galaxies build up their mass over cosmic time, and the total mass in stars per unit volume in the local universe is greater than that at higher redshift. 
	At every redshift where \textsc{UniverseMachine}\footnote{See section~\ref{subsec:models_and_constraints} for a brief description of \textsc{UniverseMachine}.} has support, we weight every galaxy in direct proportion to its corresponding stellar mass, finding that the \rev{weighted number density} of the galaxies would scale as $(1 + z)^{-0.64}$.
	
\begin{figure}[!tbp]
		\centering
		\includegraphics{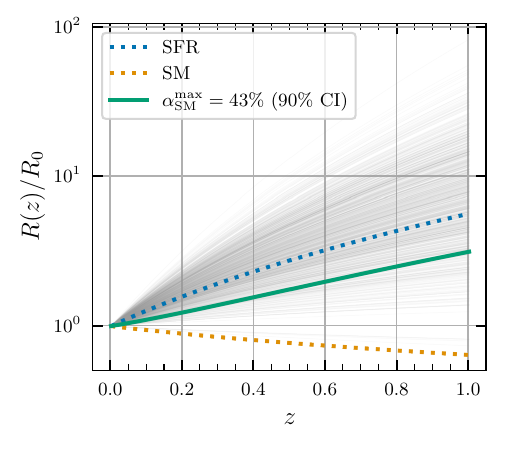}
		\caption{Redshift evolution of the merger rate inferred from GWTC-3 (black traces), plotted along with the redshift evolution of the \rev{weighted number density} of a sample of galaxies weighted by SFR (blue-dotted line) and stellar mass, $M_*$ (orange dotted line). Galaxies at higher redshifts are more star-forming than those at lower redshifts, and hence their \rev{weighted number density} follows a power law with a positive exponent: $(1 + z)^{2.5}$. On the other hand, \rev{the average galaxy stellar mass per unit volume} decreases as we go to higher redshift, and hence the \rev{$M_*$-weighted number density of galaxies} follows a power law in $(1 + z)$ with a decreasing exponent: $(1 + z)^{-0.64}$.
			The green line shows the redshift evolution with the  maximum permissible value of the mixture coefficient: $\alpha_{\rm SM}^{\rm max} = 43 \%$ at 90\% CL. }
		\label{fig:sm_sfr}
	\end{figure}

	Following~\citet{Adhikari:2020wpn}, we assume for simplicity that BBHs merge in a sample of galaxies either weighted by their stellar mass or SFR, with their relative abundances specified by a mixing fraction $\alpha_{\rm SM}$. \rev{Such a weighting in stellar mass and SFR physically corresponds to long and short delay times respectively (see Section~\ref{subsec:models_and_constraints} for a discussion).} The effective model for the rate of BBH mergers, $R(z)$, can be written as,
	\begin{equation}
		\label{eq:mixture}
		R(z) \propto \alpha_{\rm SM} (1 + z)^{-0.64} + (1 - \alpha_{\rm SM}) (1 + z )^{2.5}
	\end{equation}
	We plot the inferred $R(z)$ along with the stellar mass and SFR weighted expectations in Figure~\ref{fig:sm_sfr}.
	The black traces show the redshift evolution of the merger rate inferred from GWTC-3.
	By comparing the redshift evolution predicted by Eq.~\eqref{eq:mixture} for different values of $\alpha_{\rm SM}$ with the inference from GWTC-3, we ascertain $\alpha_{\rm SM} < 43\% \ \qty[90\% \ {\rm CL; \text{ uniform prior on } \alpha_{\rm SM}}]$ i.e. a maximum of $43\%$ of the total mergers follow a sample of galaxies weighted by stellar mass, with the rest weighted by SFR.
	The hypothesis that BBHs are hosted purely by a set of galaxies weighted by their stellar mass (i.e. $\alpha_{\rm SM} =100\%$) is ruled out by the redshift evolution of the merger rate at $> 99.9\%$.
	As a corollary, at least 57\% host-galaxies should belong to a galaxy sample weighted by SFR.
	As mentioned earlier, this model is rather simple, but it sheds light on the fact that the redshift distribution is a powerful discriminator for the set of host-galaxies within which GW events merge. For instance, in the statistical dark siren analysis to measure the Hubble parameter, individual candidate host-galaxies are typically weighted by luminosities in passbands that trace stellar mass or their star formation rate (see e.g.~\citealt{2019ApJ...871L..13F}); our results show that a stellar mass weighting is inconsistent with the redshift evolution of the BBH merger rate, and applying such a weight during the analysis should be avoided. 
	
	\section{Inferring host-galaxy properties using a physically motivated model}\label{sec:physical_constraints}
	
	We now turn to exploring a more physically-motivated model for the evolution of binaries.
	\subsection{Redshift evolution of the merger rate, and constraints on the delay-time distribution}
	\label{subsec:delay-time-constraints}

	For the purposes of this work we model the redshift evolution of the merger rate as depending on the formation rate, $R_f$, of the progenitor binaries, and the distribution of delay times, $p(t_D)$, between formation of the binary progenitors and subsequent merger of the binaries.  Assuming these, the merger rate, $R$, as a function of lookback time, $t$, can be written as
	\begin{equation}\label{eq:merger-rate}
		R(t) = \int_{0}^{\infty} \dd{t_D} R_f(t + t_D) \ p(t_D) \qq{.}
	\end{equation}
	Since redshift, $ z $, and lookback time, $ t $, are related for a given cosmology, $R(t)$ can also be equivalently written as $R(z)$. We assume that $ R_f(t) $ follows the cSFRD as inferred by \textsc{UniverseMachine} \citep{Behroozi:2019kql}\footnote{We should note that \textsc{UniverseMachine} uses a different dataset as well as a different procedure for constraining the cSFRD as compared to other works~\citep{Madau:2014bja, Madau:2016jbv}. The two results are consistent within their respective errorbars.}. For the delay-time distribution model, we employ a power-law prescription parametrized by a power-law index $\alpha$ with a minimum allowed delay-time of $ t_D^\mathrm{min} $ and maximum allowed delay of 13.5 Gyr. That is,
	\begin{equation}\label{key}
		p(t_D) = 
		\begin{cases}
			t_D^\alpha & t_D^\mathrm{min}\leq t_D\leq 13.5 \ \mathrm{Gyr} \\
			0 & \mathrm{otherwise}
		\end{cases}
	\end{equation} 
	
	\begin{figure}[!t]
		\centering
		\includegraphics{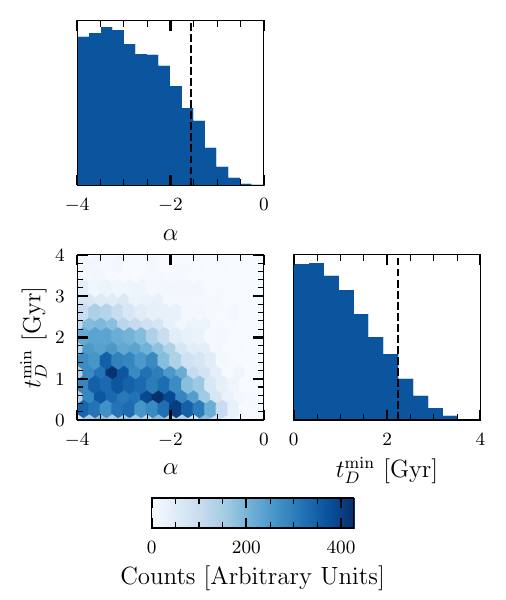}
		\caption{Constraints on the model parameters of the delay-time distribution---power law slope $\alpha$ and minimum delay-time $t^{\rm min}_D$---inferred from GWTC-3 data. GWTC-3 provides stronger constraints as compared to GWTC-2, with the $\alpha < -1.55$ and $t^{\rm min}_D < 2.23  \ {\rm Gyr}$ at 90\% CL.}
		\label{fig:delay time-constraints}
	\end{figure}
	
	\begin{figure*}[!ht]
		\centering
		\includegraphics{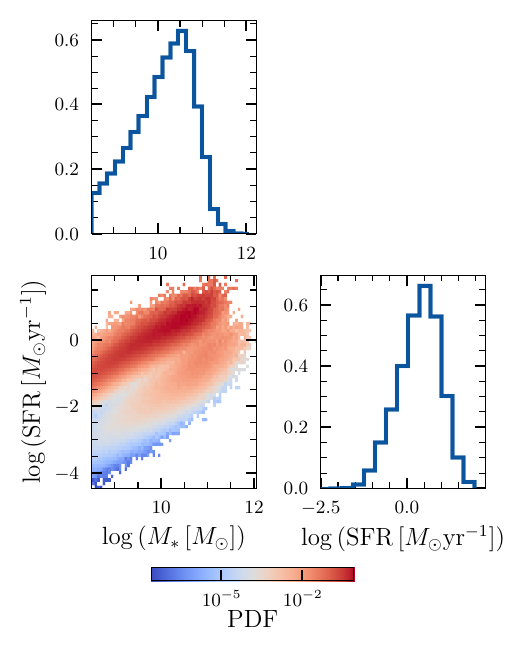}
		\includegraphics{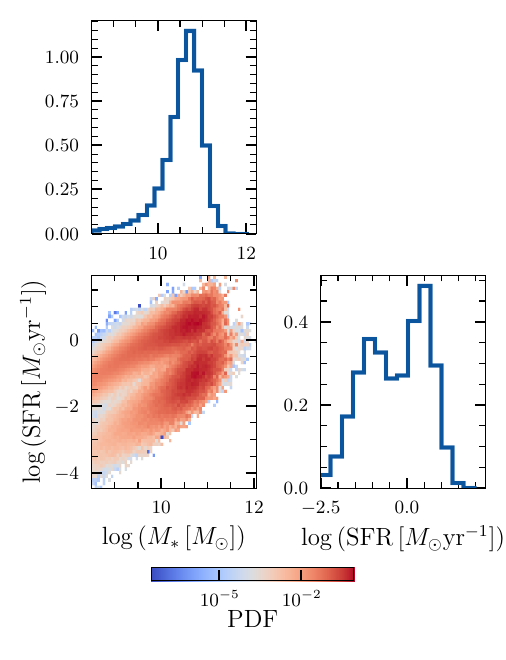}
		\caption{Two-dimensional \rev{BBH rate-weighted} histogram of host-galaxies in the $M_*$--SFR plane at $ z=0.21 $, plotted for representative short ($t_D = 0 \ {\rm Gyr}$; \textit{left}) and long ($t_D = 10 \ {\rm Gyr}$; \textit{right}) delay times. The one-dimensional histograms show the marginalized distributions of $\log \mathrm{SFR}$ and $\log M_*$. A general trend in both panels is that galaxies occupy one of two broadly distinct branches in $M_*\mbox{--}\mathrm{SFR}$ plane (visually more apparent in the right panel). The upper branch contains star forming galaxies, while the lower branch contains galaxies that have quenched their star formation. \rev{For short delay times, binaries merge quickly after forming; their host-galaxies do not evolve significantly in this duration. Hence, if binaries have short delay times, they will always track star-forming galaxies. On the other hand for long delay times between formation and merger, galaxies can evolve significantly and quench their star formation, thereby transferring some support from the star-forming branch to the quiescent branch.}}
		\label{fig:short-long-delay-times}
	\end{figure*}
	
	Latest results from the LVK collaboration \citep{2023PhRvX..13a1048A} derive constraints on the redshift evolution of the merger rate assuming that it follows $ R(z) \sim (1 + z)^\kappa $, and determine the posterior probability distribution on $ \kappa $: $p(\kappa|{\rm data})$. The LVK GWTC-3 analysis assumed
	a flat prior on $\kappa$, and so the likelihood satisfied $\mathcal{L}\qty({\rm data}| \kappa) = p\qty(\kappa | {\rm data})$. In order to compare the model expectation from Eq.~\eqref{eq:merger-rate} to redshift evolution inference, we need to map the delay-time distribution parameters onto a prediction on $\kappa$. Let us denote this model prediction by $\tilde{\kappa}\qty(\alpha, t_D^\mathrm{min})$. Due to the functional form chosen for $R(z)$, $\tilde{\kappa}$ can be approximated as
	\begin{equation}
		\label{eq:approxkappa}
		\tilde{\kappa}\qty(\alpha, t_D^\mathrm{min}) = \log_2\qty[\dfrac{R(z=1; \alpha, t_D^\mathrm{min})}{R(z=0; \alpha, t_D^\mathrm{min})} ] \qq{.}
	\end{equation}
	Since current observations measure the merger rate only out to $ z=1 $, this is a good approximation; however, with more sensitive detectors, the parameters of the delay-time distribution would have to be constrained directly by jointly sampling over them along with hyperparameters of the underlying population model related to mass, spin, etc\footnote{\rev{Recent work~\citep{2023ApJ...946...16E,2024PhRvX..14b1005C} has found hints of structure in the redshift evolution of the BBH merger rate beyond the $(1 + z)^\kappa$ model we assume here, and this model mis-specification could impact our results. However, since the data are still consistent with the $(1 + z)^\kappa$ model, we ignore the effects of these deviations.}}. Using the approximation in Eq.~\eqref{eq:approxkappa}, we can write a likelihood function for $\alpha$ and $t^{\rm min}_D$,
	\begin{equation} 
		\mathcal{L}\qty({\rm data} | \alpha, t_D^\mathrm{min}) = \mathcal{L}\qty({\rm data} | \tilde{\kappa}\qty(\alpha,t_D^\mathrm{min} )) \qq{,}
	\end{equation}
	where the RHS is given by the LVK GWTC-3 population analysis.
	
	We construct the joint posterior on $\alpha$ and $t_D^\mathrm{min}$ using the likelihood function above. For this, we assume flat priors on $\alpha$ betweem $\qty[-4,1]$, and on $t_D^\mathrm{min}$ between $[0, 6] \ {\rm Gyr}$.
	We use the Markov chain Monte Carlo (MCMC) sampler \texttt{emcee}~\citep{2013PASP..125..306F} to sample from the posterior distribution. Constraints on $\alpha$ and $t^{\rm min}_D$ from GWTC-3  are shown in Figure \ref{fig:delay time-constraints}. As expected, the constraints on the parameters are stronger than those derived from GWTC-2~\citep{2021ApJ...914L..30F} due to the {50\% additional} events included in GWTC-3. We constrain $\alpha < -1.55 $ and $t_D^{ \rm min} < 2.23 \ {\rm Gyr}$ at 90\%~CL. 
	On the whole, the posteriors for both parameters rail against the lower limit of the prior, illustrating a preference for {short} delay times in the population. This is consistent with expectations that BBHs follow the cSFRD, and also with other recent work~\citep{2021ApJ...914L..30F,2023arXiv230715824F,2023arXiv231017625T,2023MNRAS.523.4539K}. 
	
	\subsection{Modelling galaxy formation and evolution, and constraining host-galaxy distributions}
	\label{subsec:models_and_constraints}
	\begin{figure*}[!ht]
		\centering
		\includegraphics{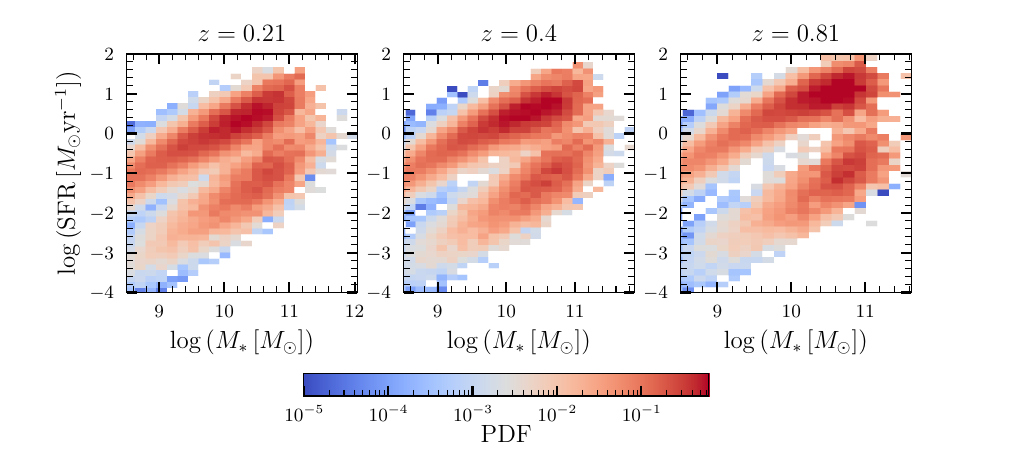}
		\caption{Median inferred \rev{BBH rate-weighted} host-galaxy distribution in the $\log M_*$--$\log \mathrm{SFR}$ plane for three different redshifts $z=0.21, 0.4, 0.81$. These specific values of redshift are chosen for illustrative purposes; our model can predict the host-galaxy distribution at any redshift at which a simulation snapshot exists. At all three redshifts, the distribution has support both in the star-forming branch as well as the quiescent branch, although it has more support in the former. This is consistent with the fact that the delay-time distribution has support for non-zero delay times.}
		\label{fig:inferred-hosts}
	\end{figure*}
	
	To obtain observed galaxy properties across cosmic time, we use the \textsc{UniverseMachine}~\citep{Behroozi:2019kql} set of semi-analytical galaxy formation simulations. Instead of directly simulating the gas and star-formation during galaxy formation as is done in hydrodynamical simulations such as IllustrisTNG~\citep{2018MNRAS.473.4077P, 2019ComAC...6....2N} or EAGLE~\citep{2015MNRAS.450.1937C}, \textsc{UniverseMachine} populates galaxies within dark matter halos in a pure cold dark matter simulation with a semi-empirical model. This model relates the star-formation rate to the halo mass accretion rate, and constrains this with a Monte Carlo scheme using a wide variety of observations  over a range of redshifts out to $z=10$. More details about \textsc{UniverseMachine} can be found in \citet{Behroozi:2019kql}. As a result, one can track the evolution of galaxy properties (e.g.  stellar mass, host halo mass, star formation history) across redshift.
	We use publicly available \textsc{UniverseMachine} mock catalogs\footnote{\href{https://halos.as.arizona.edu/UniverseMachine/DR1/SFR/}{https://halos.as.arizona.edu/UniverseMachine/DR1/SFR/}} created using the Bolshoi-Planck dark matter only simulations \citep{2011ApJ...740..102K} with a box size of $250\,{h}^{-1} {\rm Mpc}$ and $2048^3$ particles. These catalogs allow us to extract galaxy properties in $170$ logarithmically-spaced redshift bins between $z=0$ to $z=13$.  
	While we use \textsc{UniverseMachine} for results in this work, the procedure can be repeated with any galaxy formation simulation that tracks galaxy properties across redshift.

	For each galaxy at redshift $z_0$, we calculate the merger rate as
	\begin{equation}
		R^{\rm merg}_i(z_0) = \int_{}^{} \dd{t_D} R_i^{\rm SFH}(t(z_0) + t_D) \times p(t_D) \qq{,}
	\end{equation}
	where $R^{\rm merg}_i(z)$ is merger rate of the $i$'th galaxy at redshift $z$, while $R_i^{\rm SFH}(t(z))$ is the star-formation history (SFH) of the $i$'th galaxy numerically interpolated across redshift bins.
	The equation above is the same as Eq.~\eqref{eq:merger-rate}, except that the quantities pertain to a specific galaxy rather than the total merger rate across all galaxies. 
	When delay times are short, the merger rate of BBHs will have a redshift evolution very close to that of the SFR density, since in our model the rate of production of binaries is directly proportional to the rate of production of stars.
	On the other hand when delay times are long, the galaxies would have evolved significantly, and would hence no longer necessarily be star-forming at the merger epoch. These quiescent galaxies would have a very different evolution for their \rev{weighted number density} as a function of redshift, which would be more similar to a sample of host-galaxies weighted by their stellar mass rather than their SFR.
	
	We illustrate this phenomenon in Fig.~\ref{fig:short-long-delay-times}, where we plot the two-dimensional \rev{BBH rate-weighted histogram} of the stellar mass and SFR of the $z=0.21$ host-galaxies assuming $p(t_D| t_0) = \delta(t_D - t_0)$. In this $M_*$--SFR plane, the distribution of galaxies has two broadly distinct branches; the upper branch corresponding to star-forming galaxies, and the lower branch corresponding to quiescent galaxies. Within these two branches $M_*$ and SFR are, on an average, correlated positively. We choose two representative values, $t_0 = 0\,{\rm Gyr}$ and $10\,{\rm Gyr}$, corresponding to short and long delay times, respectively. When delay times are short, the star forming branch is highly populated, while the quiescent branch is relatively empty---this is  because binaries merge at roughly the same redshift that they formed, and their host-galaxies do not have much time to quench their star formation (hence stay star-forming). Naturally, the redshift evolution of the \rev{weighted number density} in such a scenario follows the SFR. This scenario changes when delay times are large---the quiescent branch is now more populated since galaxies have had a long time to evolve and quench their star formation. The level to which the quiescent branch is populated increases with an increase in the delay time.
	
	\begin{figure*}[!ht]
		\centering
		\includegraphics{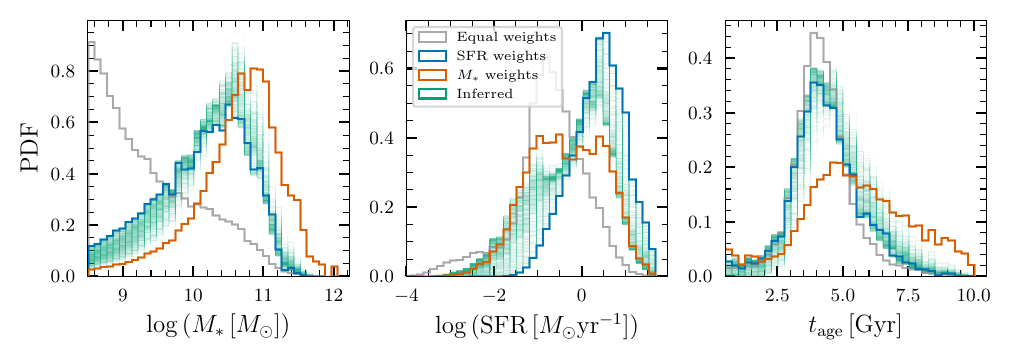}
		\caption{Inferred \rev{rate-weighted} histograms (green) on $M_*$, SFR, and mean stellar age $t_\mathrm{age}$ \rev{of BBH host-galaxies} at $z=0.21$, plotted along with histograms for samples weighted equally (grey), by SFR (blue), and by $M_*$ (orange). For all the quantities, the inferred curves are close to the SFR-weighted curves, with the scatter being due to the non-zero delay times.}
		\label{fig:sm_sfr_age}
	\end{figure*}
	
	\subsection{Results}
	
	In order to extract host-galaxy properties from the observed redshift evolution of the merger rate, we convolve the constraints on the delay-time distribution model obtained in Section~\ref{subsec:delay-time-constraints} with galaxy SFHs taken from \textsc{UniverseMachine}. For each sample drawn from the joint posterior on $\alpha$ and $t_D^{\rm min}$, we track every galaxy in the catalog and assign to each one a merger rate of binaries according to Eq.~\eqref{eq:merger-rate} at every epoch. 
	We obtain a set of mock candidate BBH host-galaxies, with complete information about their current stellar mass and SFR, mass assembly and star formation history, as well as merger rate history. Since \textsc{UniverseMachine} models the connection between galaxies and their host halos, we also obtain information about their host halo mass, radius, and peculiar velocity.
	\rev{Corresponding to every posterior sample, one can plot a histogram of galaxy properties weighted appropriately by the merger rate of BBHs in that galaxy; this {\textit{BBH rate-weighted histogram}} will indicate the population of galaxies hosting BBH events}.

	\subsubsection{Stellar mass, star formation rate, and mean stellar age}
	
	Fig.~\ref{fig:inferred-hosts} shows the median inferred distribution of BBH host-galaxies inferred in the $M_*\mbox{--}{\rm SFR}$ plane using our prescription at redshifts $z = 0.21, 0.4, 0.81$. One noticeable broad feature is that the star-forming branch is shifted to higher values of SFR at high redshifts. This is because, consistent with expectations, galaxies at higher redshift on an average form more stars. Notably, there is significant support for BBH host-galaxies both in the star-forming branch as well as the quiescent branch, although there is more support in the star-forming branch. This is a direct result of the small but nonzero delay times that are allowed by the constraints in Section~\ref{subsec:delay-time-constraints}---these delay times are large enough to induce some support for galaxies in the quiescent branch, but not so large as to entirely populate this branch.
	The median stellar mass of the hosts at $z=0.21$ is $1.9 \times 10^{10}\,M_\odot$, while the median SFR is $0.9\,M_\odot \mathrm{yr}^{-1}$. The median stellar mass at $z=0.4$ and $z= 0.81$ is $2.1 \times 10^{10}\,M_\odot$ and $2.1 \times 10^{10}\,M_\odot$ respectively, while the median SFR is $1.6\,M_\odot \mathrm{yr}^{-1}$ and $4.5\,M_\odot \mathrm{yr}^{-1}$ respectively.
	
	Fig.~\ref{fig:sm_sfr_age} shows the one-dimensional \rev{BBH rate-weighted histograms} on $M_*$, SFR, and mean stellar age, $t_{\rm age}$, inferred from our prescription at $z=0.21$. Here, $t_{\rm age}$ is defined on a per galaxy level as,
	\begin{equation}
		t_{\rm age} = \dfrac{\sum_i t_i \times R^{\rm SFH}(t_i) }{\sum_i  R^{\rm SFH}(t_i)}\qq{,}  t_i = t(z_i) - t(z_0)\qq{.}
	\end{equation}
	We also plot, for comparison, histograms for each of the parameters weighted equally, by $M_*$, and by SFR. The equal weights case should be thought of as the set of all galaxies in the Universe at a given redshift, while the $M_*$-weighted and the SFR-weighted galaxies track massive and star-forming galaxies respectively. The inferred histograms are plotted in green, and the width of the green region should be thought of as coming from the uncertainties in measuring the parameters of the delay-time distribution
	
	We see that the distribution of $M_*$ is broadly consistent with the SFR-weighted sample of galaxies, and is inconsistent with an $M_*$-weighted sample of galaxies following expectations from Fig.~\ref{fig:sm_sfr}. Most of the inferred histograms cluster around the SFR-weighted histograms, consistent with the railing of $\alpha$ and $t_D^{\rm min}$ towards the lower bound on the prior in Fig.~\ref{fig:delay time-constraints}.
	A similar trend can be seen in the distribution of SFRs as well, which is consistent with an SFR-weighted sample of galaxies, but also has support for galaxies with slightly lower SFRs. 
	The inferred mean stellar age of host-galaxies is $4.4  \,{\rm Gyr}$, consistent with that of star-forming galaxies.

	The general picture at $z=0.21$ also holds at the higher redshifts which we considered, albeit with larger errors, since the merger rate at higher redshifts is less well-constrained. We discuss results for $z=0.4, 0.81$ briefly in Appendix~\ref{sec:appendix}.

	\subsubsection{Magnitudes}
	
	The light that we observe from galaxies encodes signatures of their star formation and mass assembly history. Given information about the initial mass function and star formation history of galaxies, one can forward-model galaxy spectra using stellar population synthesis models (see e.g.~\citealt{2003MNRAS.344.1000B} and \citealt{2009ApJ...699..486C}). These spectra can then be bandpassed appropriately to derive luminosities and magnitudes. For our inferred host-galaxies, we derive (absolute) magnitudes in the Johnson-Morgan B-band~\citep{1953ApJ...117..313J} and the 2MASS K$_s$-band~\citep{2006AJ....131.1163S}\footnote{The passbands are downloaded from the SVO Filter Profile Service \citep{2020sea..confE.182R, 2012ivoa.rept.1015R}: \href{http://svo2.cab.inta-csic.es/theory/fps/index.php}{http://svo2.cab.inta-csic.es/theory/fps/index.php}.}, thought to be proxies for the SFR and stellar mass, respectively~\citep{2003ApJS..149..289B,2016ApJ...829L..15S}. To simulate mock galaxy spectra, we use the stellar population synthesis model from \citet{2003MNRAS.344.1000B} implemented in the \texttt{bagpipes} software package~\citep{2018MNRAS.480.4379C}. Since we already have simulated star formation histories from \textsc{UniverseMachine} for our mock galaxies, we use the ``custom'' SFH implemented in  \texttt{bagpipes} instead of relying on semi-analytical prescriptions \rev{for the SFH}. Also, since \textsc{UniverseMachine} does not model metallicity evolution, we fix the metallicity $Z=0.5\,Z_\odot$ at all redshifts, where $Z_\odot=0.02$ is solar metallicity\footnote{We have checked that varying $Z$ between $10^{-3}\,Z_\odot$ and $3\, Z_\odot$ only changes the calculated magnitude by at most 0.75 units. The shape of the histogram of magnitudes stays approximately the same, with a constant shift to the left or the right. }. 
	
	\begin{figure}
		\centering
		
		\includegraphics{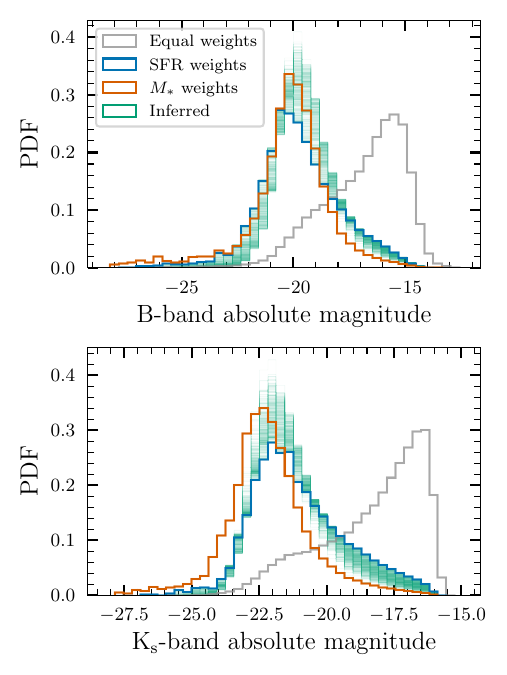}
		
		\caption{\rev{Rate-weighted histogram} of inferred absolute magnitudes of BBH host-galaxies in the B-band and the K$_s$-band (commonly used as proxies for the SFR and stellar mass, respectively) at $z=0.21$.}
		\label{fig:mag}
	\end{figure}
	
	As before, we plot the \rev{BBH rate-weighted histogram} on the inferred magnitudes in the B-band and K$_{\rm s}$-band, along with histograms weighted equally, by $M_*$ and by SFR, in Fig.~\ref{fig:mag}. In both bands, the equally weighted set of galaxies has a preference for high magnitudes (i.e. low luminosities), since there are significantly more low luminosity galaxies in the Universe (see for e.g.~\citealt{1976ApJ...203..297S}). In contrast, the histogram for the inferred weights peaks at lower magnitudes (i.e. higher luminosities). The median inferred magnitude in the B-band is {$-19.6$}, while that in the  K$_{\rm s}$-band is {$-21.5$}. 
	These inferred histograms on the magnitudes can be directly used to calculate luminosity-weights in the dark-siren method for measuring the cosmic expansion history with GW events. {In addition, they may enable the design of more optimized follow-up strategies to search for possible electromagnetic counterparts to BBH mergers.}

	\begin{figure*}[!htbp]
		\centering
		
		\includegraphics{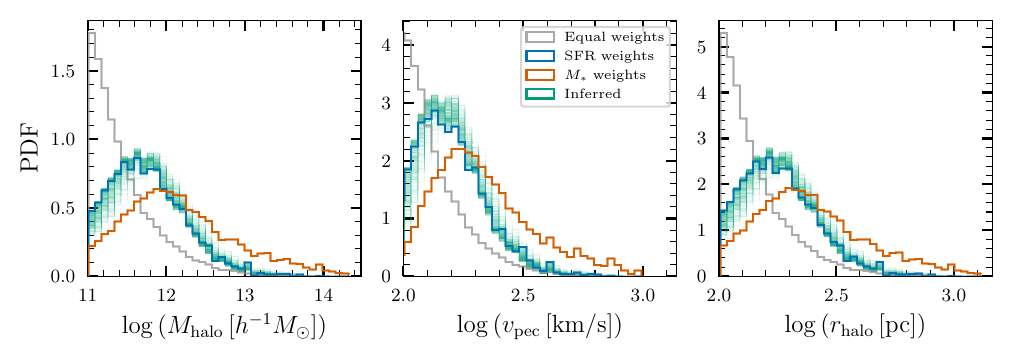}
		\caption{Inferred distributions of the host dark matter halo mass $M_{\rm halo}$, host peculiar velocity $v_{\rm pec}$, and halo radius $r_{\rm halo}$ at $z=0.21$}
		\label{fig:m_r_v}
	\end{figure*}
	
	\subsubsection{Halo mass, halo radius, and peculiar velocity}
	The formation and evolution of galaxies is intrinsically connected to their host dark matter halos (see \citealt{2018ARA&A..56..435W} for a review); thus, the properties of a galaxy are correlated with its parent dark matter halo\footnote{This correlation is referred to as the ``galaxy-halo'' connection.}. Specifically, larger halos host larger and more-massive galaxies. This correlation also naturally causes massive galaxies to have a greater peculiar velocity, due to the denser environment (see e.g.~\citealt{2001MNRAS.322..901S}).
	
	Since \textsc{UniverseMachine} also tracks host halo properties along with the galaxy properties, we also have host halo information for the host-galaxies. Fig.~\ref{fig:m_r_v} shows the \rev{rate-weighted histograms} on inferred host dark matter halo mass, $M_{\rm halo}$, halo radius, $r_{\rm halo}$, and galaxy peculiar velocity, $v_{\rm pec}$. We note that $M_*$-weighted galaxies have higher $M_{\rm halo}$, $v_{\rm pec}$, {and $r_{\rm halo}$}, as compared to those weighted by SFR; this is because the former galaxies are larger and hence lie in larger dark matter halos. Similar to the case of galaxies presented in the previous sub-sections, we find that the inferred histograms for host dark matter halos lie closest to the halos associated with SFR-weighted galaxies. The median $M_{\rm halo}$ and $r_{\rm halo}$ are {$5.4 \times 10^{11}\,h^{-1}M_\odot$} and {$180\,{\rm pc}$}, respectively.
	The median $v_{\rm pec}$ is {$150 \,{\rm km / s}$}, but has a tail extending out to {$v_{\rm pec}\approx 800 \ {\rm km / s}$}. This means that the peculiar velocities can shift the redshifts of BBHs at most by {$v_{\rm pec} / c \approx 0.003$}, and thus are unlikely to affect the uncertain estimates of distances derived from BBHs even in next-generation detectors.
	
	\subsubsection{Large-scale bias}
	
	Galaxies are \textit{biased} tracers of the underlying matter density in the Universe---they prefer forming at rare, highly-dense peaks of the underlying matter density field. \citet{1984ApJ...284L...9K} showed that the clustering of galaxies, quantified by the two-point correlation $\xi_{\rm gal}(r)$, can be modelled as a constant enhancement over the clustering of the underlying matter distribution on large-scales: $\xi_{\rm gal}(r) = b^2_{\rm gal} \xi_{\rm m}(r)$. \rev{The value of the bias}, $b_{\rm gal}$, is directly related to the masses of halos that host-galaxies (see e.g. \citealt{2002PhR...372....1C}). 
	
	\begin{figure}[!htbp]
		
		\centering
		\includegraphics{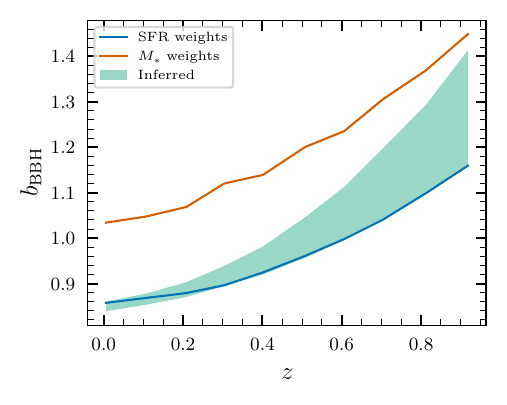}
		
		\caption{Large-scale bias of BBHs, $b_{\rm BBH} $, as a function of redshift, $z$, as predicted by the host properties inferred in this work. }
		\label{fig:bias}
	\end{figure}
	
	Since BBHs also form and evolve in galaxies, they are a biased tracer of the underlying matter density. We denote $b_{\rm BBH}$ as the large-scale bias of BBHs, akin to the same quantity for galaxies.
	Indeed, multiple works have proposed using the bias to probe the astrophysical formation channels of GW transients~\citep{2016PhRvD..94b4013N,2016PhRvD..94b3516R,2018JCAP...09..039S, 2021PhRvD.103d3520M,2020arXiv200501111V}.
	Using the halo mass-bias relation of \citet{2010ApJ...724..878T}, implemented in the \texttt{colossus} software package~\citep{2018ApJS..239...35D}, we are able to predict values for the large-scale bias {of BBHs} from the inferred host-galaxy distributions.
	
	We plot the 90\% credible region of the inferred \rev{BBH rate-weighted bias} as a function of redshift in Fig.~\ref{fig:bias}. Overall, the bias has an increasing trend towards high redshift---this is a natural consequence of hierarchical structure formation in the Universe. We note that the bias at $ z=0 $ is predicted to be in the range {$0.86\mbox{--}0.88$ at $z=0.1$ at 90\% CL, while that at $ z=0.9 $ is predicted to be $1.16\mbox{--}1.40$ (90\% CL)}. \rev{The inferred bias is consistent with that of SFR-weighted galaxies, but is inconsistent with $M_*$-weighted galaxies; this is because the bias is purely a function of $M_{\rm halo}$, and the inferred $M_{\rm halo}$ distribution is inconsistent with $M_*$-weighted galaxies (see Fig.~\ref{fig:m_r_v})}.
	
	\section{Summary and Outlook}
	\label{sec:summary}
	We have shown that the measured redshift evolution of the BBH merger rate from GWTC-3 constrains the set of host-galaxies that BBHs are associated with. We started with a simple mixture model of galaxies weighted either by SFR or stellar mass. {We demonstrated that the hosts of merging BBHs do not exclusively track stellar mass} at the {$> 99.9\%$} level---at most 43\% of host-galaxies come from a stellar mass weighted sample. We then used \textsc{UniverseMachine} simulations, along with data-driven constraints on the BBH delay-time distribution, to infer the host-galaxy population. Our results indicate that most BBHs are hosted by actively star-forming galaxies, but there is support for a small fraction of them being hosted by quiescent galaxies. We provide inferred distributions for the halo masses, radii, peculiar velocities, and stellar ages. {We also make predictions for the observed ${\rm K_s}$- and B-band magnitudes of BBH host-galaxies, as well as the large-scale bias associated with the clustering of BBHs.} \rev{Our results assume that the $R(z) \sim (1+z)^\kappa$ is a good description for the redshift evolution of the merger rate. There are some hints of structure in the $R(z)$ beyond this power-law prescription~\citep{2023ApJ...946...16E, 2024PhRvX..14b1005C}; if these results hold up with newer data, it might necessitate invoking the presence of two (or more) populations of BBHs with distinct delay-time distributions. However, the inferred delay-time distribution would still prefer short delay-times, and hence we do not expect our results to change significantly.}
	
	Unlike previous works which provide distributions of host-galaxies based on prescriptions from population synthesis, our framework uses constraints from {{\em observed}\/ GW data to solve the inverse problem: we infer a set of host-galaxies without assumptions for any specific prescription for binary formation and evolution.}
	While we have concentrated on BBHs in this work, our framework will extend trivially to BNSs and NSBHs, once we are able to measure the redshift evolution of their merger rates.  This framework can also be extended to infer the host-galaxy population of astronomical objects for which source localization is poor (e.g. poorly localized gamma-ray bursts and fast radio bursts). Our inferred distributions on various parameters can be used  to design optimal weighting schemes for the dark-siren method to constrain the expansion history of the Universe. In addition, our predictions for host-galaxy properties may aid in the design of optimal follow-up strategies for BBH events. {This will be of particular interest once the BNS and NSBH extensions to our methods have been implemented, since those searches are expected to have transient electromagnetic counterparts.}
	
	For the main results in this work, we have assumed that binaries followed a physically motivated merger-rate evolution, given by the \textsc{UniverseMachine} cSFRD convolved with a delay-time distribution. This prescription is expected to work well if all binaries form and merge in isolated galactic field environments, \rev{and our results should be interpreted in this context. For instance, binaries could also form and merge in dynamical environments such as globular clusters (GCs) or active galactic nuclei (AGN), and these populations will have different progenitor formation rates and delay-time distributions, thus requiring changes to our modeling procedure. Our framework can be readily extended to include these channels given some assumption of the progenitor formation rate, although adding new parameters is likely to increase the statistical error on the results.}  \rev{Since compact binary formation is expected to be a strong function of metallicity~\citep[and references therein]{2024AnP...53600170C}, another natural extension of our framework would be to include the metallicity-specific star-formation history from hydrodynamical simulations of galaxy and structure formation while assigning merger rates to individual galaxies. This would also necessitate using a metallicity-specific cSFRD as the progenitor formation rate in constraining delay-time distributions. It is also possible to relax our assumption of specific progenitor formation rates and delay-time distributions, and instead generalize our approach to compare redshift evolution of galaxies and BBHs in a non-parametric fashion, although this would still be subject to any modeling differences inherent in different galaxy formation models}.
	
	We have demonstrated the power of {\em observed}\/ properties of the existing GW population to constrain properties of the associated host-galaxies. The relation between GW mergers and their parent galaxies offers a rich vein to mine for astrophysical insights.
	A more precise future measurement of the rate evolution of binaries will help us further constrain the host environments, possibly helping to disentangle the various formation channels that contribute to the total population of merging binaries observable with ground-based GW detectors.
	
	\section*{Acknowledgments}
	We are grateful to Jacqueline Antwi-Danso for patiently answering all our questions about galaxy SEDs, and Peter Behroozi for help with interpreting the \textsc{UniverseMachine} catalogs. We also thank Surhud More for useful discussions, and Alexandra Hanselman and Aditya Kumar Sharma for a reading of our draft. 
	All computations for this work were performed on the Alice cluster at ICTS-TIFR. 
	
	AV's work is supported by the Department of Atomic Energy, Government of India, under Project No. RTI4001. AV also acknowledges support by a Fulbright Program grant under the Fulbright-Nehru Doctoral Research Fellowship, sponsored by the Bureau of Educational and Cultural Affairs of the United States Department of State and administered by the Institute of International Education and the United States-India Educational Foundation. AV and MF are supported by the Natural Sciences and Engineering Research Council of Canada (NSERC) (funding reference number 568580). 
	D.E.H is supported by NSF grants AST-2006645 and PHY-2110507, as well as by the Kavli Institute for Cosmological Physics (KICP) through an endowment from the Kavli Foundation and its founder Fred Kavli.
	
	We thank KICP for support through ``The quest for precision gravitational wave cosmology'' workshop where this work was initiated, and ICTS for support through the Largest Cosmological Surveys and Big Data Science (code: ICTS/BigDataCosmo2023/5) workshop. This research has made use of the Spanish Virtual Observatory (https://svo.cab.inta-csic.es) project funded by MCIN/AEI/10.13039/501100011033/ through grant PID2020-112949GB-I00.
	
	This material is based upon work supported by NSF's LIGO Laboratory which is a major facility fully funded by the National Science Foundation.
	
	\software{numpy~\citep{2020Natur.585..357H}, scipy~\citep{2020NatMe..17..261V}, matplotlib~\citep{2007CSE.....9...90H}, astropy \citep{2013A&A...558A..33A,2018AJ....156..123A}, jupyter~\citep{2016ppap.book...87K}, pandas~\citep{mckinney-proc-scipy-2010}, seaborn~\citep{2021JOSS....6.3021W}, bagpipes~\citep{2018MNRAS.480.4379C}, colossus~\citep{2018ApJS..239...35D} }

	\bibliography{references}
	\bibliographystyle{aasjournal}
	
	\appendix
	\section{Inferred host-galaxy properties at redshifts of 0.4 and 0.81}
	\label{sec:appendix}
\begin{figure}[!htbp]
		\centering
		\includegraphics{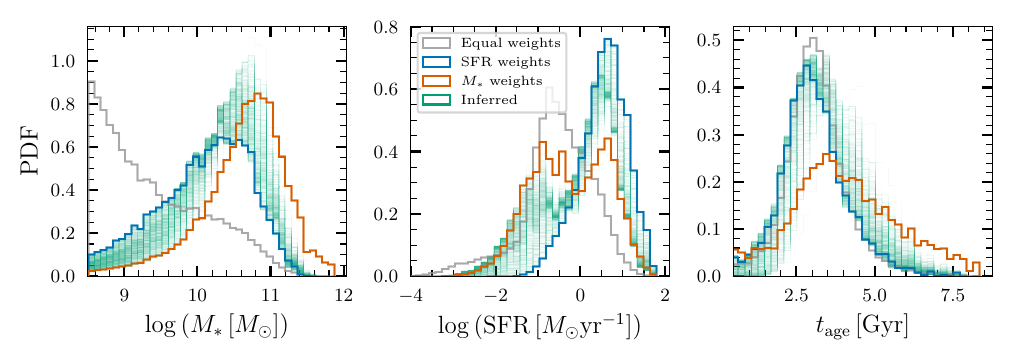}
		\includegraphics{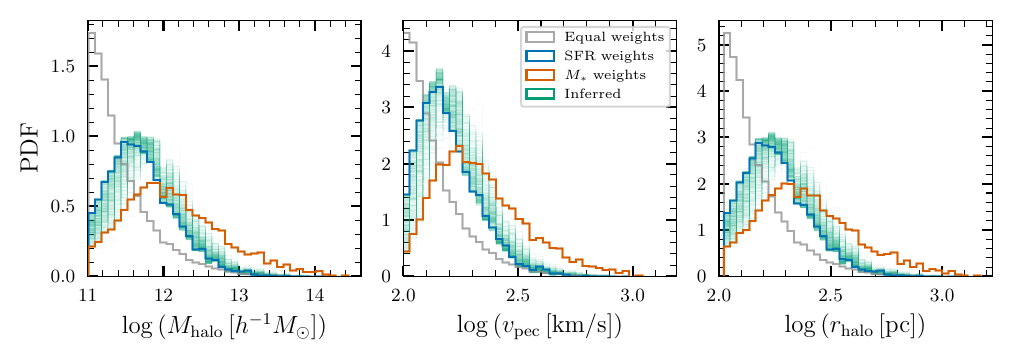}
		\caption{host-galaxy properties at $z=0.40$.}
		\label{fig:propertiesz0.4}
	\end{figure}

\begin{figure}[!htbp]
		\centering
		\includegraphics{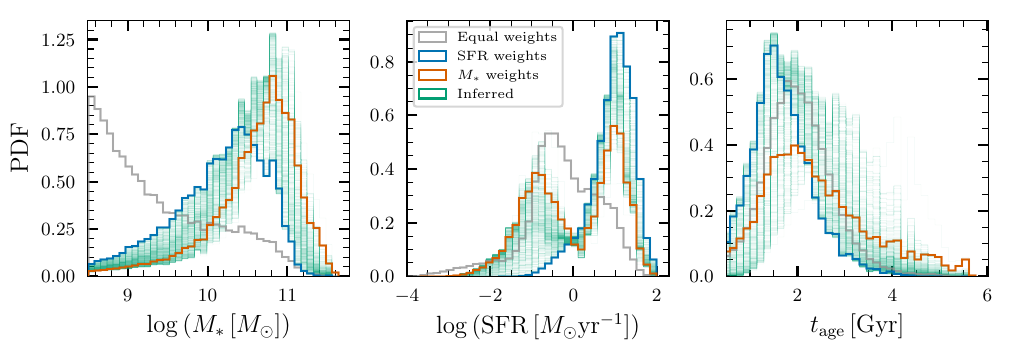}
		\includegraphics{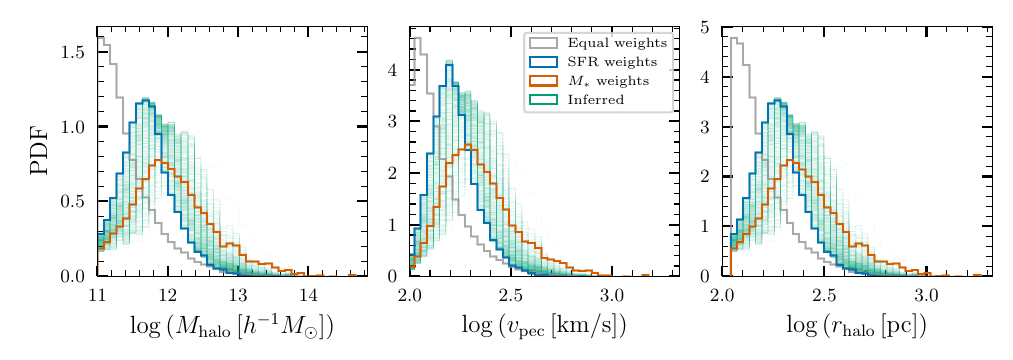}
		\caption{host-galaxy properties at $z=0.81$.}
		\label{fig:propertiesz0.81}
	\end{figure}

\begin{figure}[!htbp]
		\centering
		\includegraphics{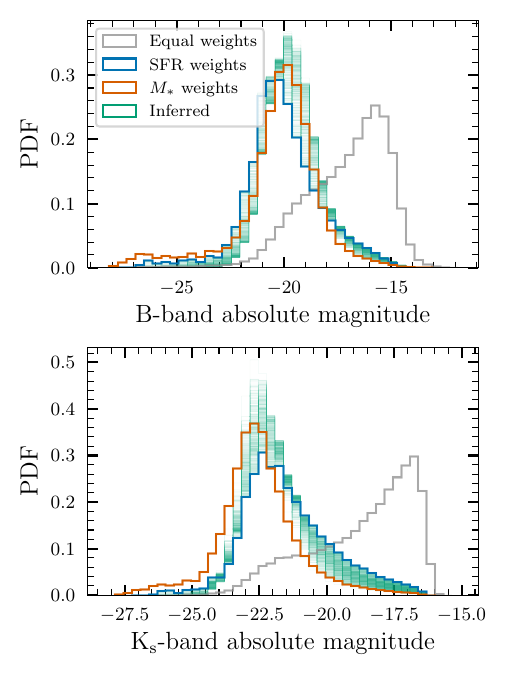}	\includegraphics{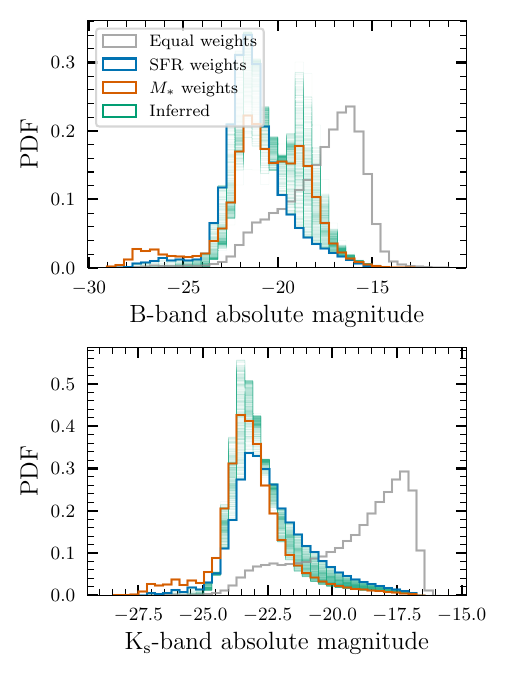}
		\caption{B-band and K$_s$-band absolute magnitudes inferred at $z=0.40$ (left) and  $z=0.81$ (right).}
		\label{fig:mags_z}
	\end{figure}

	In this section we provide inferred \rev{rate-weighted histograms} of the $M_*$, SFR, $t_{\rm age}$, $M_{\rm halo}$, $r_{\rm halo}$, and $v_{\rm pec}$, at redshifts of 0.40 (Fig.~\ref{fig:propertiesz0.4}) and 0.81 (Fig.~\ref{fig:propertiesz0.81}). We also show inferred histograms of B-band and K$_{\rm s}$-band absolute magnitudes in Fig.~\ref{fig:mags_z}. We reiterate that our prescription is able to infer galaxy properties at any redshift where a \textsc{UniverseMachine} simulation snapshot exists.
	
\end{document}